\documentclass[fleqn,usenatbib,useAMS]{mnras}
\usepackage{graphicx}
\usepackage{bm,amsmath,amssymb, ulem}
\newcommand{\pd}[2]{\frac{\partial {#1}}{\partial {#2}}}
\newcommand{\od}[2]{\frac{\mathrm{d} {#1}}{\mathrm{d} {#2}}}

\title[Biermann battery induced by streaming cosmic rays]{Biermann Battery Powered by Resistive Heating Induced by Cosmic Ray Streaming}
\author[S. L. Yokoyama and Y. Ohira]{Shota L. Yokoyama $^{1}$
    \thanks{E-mail: \href{mailto:s_yokoyama@eps.s.u-tokyo.ac.jp}{s\_yokoyama@eps.s.u-tokyo.ac.jp}},
    Yutaka Ohira $^{1}$
    \\
    $^{1}$ Department of Earth and Planetary Science, The University of Tokyo, 7-3-1 Hongo, Bunkyo-ku, Tokyo 113-0033, Japan
}

\date{Accepted XXX. Received YYY; in original form ZZZ}
\pubyear{2022}

\begin{document}
\label{firstpage}
\pagerange{\pageref{firstpage}--\pageref{lastpage}}
\maketitle

\begin{abstract}
    It is recently proposed that cosmic rays generate a seed magnetic field in the early universe.
    In this paper, we propose another generation mechanism of magnetic fields by cosmic rays, which is the Biermann battery driven by resistive heating induced by the streaming of cosmic rays. 
    This mechanism is dominant in small-scale, low-temperature, and strongly-ionized regions, compared with other previously proposed mechanisms.
    Because cosmic rays are expected to be accelerated after the death of the first stars, this mechanism can work during structure formation in the early universe.
    We show that it makes the seed magnetic field with sufficient strength for the subsequent dynamo to amplify it to the micro Gauss level in the current galaxies.
\end{abstract}


\section{Introduction}
\label{sec:1}
The existence and importance of magnetic fields and cosmic rays (CRs) are widely recognized in the current universe.
Energy densities of magnetic fields and CRs in the local interstellar medium are comparable to those of other components including thermal particles, starlight, and turbulence \citep[e.g.][]{Draine2011}.
Magnetic fields are observed on various scales from the planetary scales to the scale of galaxy clusters.
The magnetic field greater than $10^{-16} \, \mathrm{G}$ is inferred even in void regions of large scale structures from gamma-ray observation of $\mathrm{TeV}$ blazers \citep[][]{Neronov2010a}.
However, the origin of these magnetic fields is not revealed yet.

The origin of CRs is also a mystery.
Recently, \citet{Ohira2019} discussed that supernova remnants of the first stars can produce collisionless shocks and accelerate CRs by diffusive shock acceleration \citep[][]{Krymsky1977, Axford1977, Bell1978, Blandford1978}.
Because the first explosions are thought to occur at redshift $z \sim 20$, CRs are expected to exist at that time.

Magnetic fields of $\mathrm{\mu G}$ level are commonly observed in galaxies \citep[][]{Han2017, Akahori2018}.
The widely accepted paradigm to generate this strength of magnetic fields is to amplify weaker magnetic fields by some kinds of dynamo processes \citep[e.g.][]{Kulsrud1997, Kulsrud2008}.
This dynamo amplification requires the existence of "seed" magnetic fields.
There are roughly two classes for generation of seed magnetic fields, that is, primordial and astrophysical generation.
Primordial origins of magnetic fields are intensively investigated \citep[see reviews for e.g.][]{Widrow2002, Widrow2012, Durrer2013, Subramanian2016, Vachaspati2021}.
In this scenario, magnetic fields are generated during the inflation or during the electroweak or QCD phase transitions.
Recently, \citet{Mtchedlidze2022} investigated the evolution of primordially generated magnetic fields during structure formation by using MHD simulations.

Just after the beginning of structure formation, astrophysical processes become responsible for magnetic field generation.
The Biermann battery is the most commonly discussed mechanism \citep{Biermann1950}.
Because it needs a special pressure structure of electrons to drive an electron vortex, various mechanisms to make such a special structure are proposed.
For example, \citet{Hanayama2005} argued that the Biermann battery can be driven in the supernova explosions of the first stars, while \citet{Shiromoto2014} considered {the Biermann battery} in the accretion disks around proto-first-stars.
\citet{Kulsrud1997} demonstrated by numerical simulations and analytical calculations that the Biermann battery mechanism works at the shocks of forming galaxies and the generated magnetic fields can be further amplified by dynamo processes.
Magnetic fields can also be generated in the epoch between the phase transitions and the structure formation.
\citet{Ichiki2006} discussed that second-order coupling between photons and electrons causes magnetogenesis before the recombination era.
\citet{Naoz2013} proposed that, after the recombination, the Biermann battery is driven by a baroclinic flow of residual free electrons.

Based on the expectation that the first CRs exist at $z\sim 20$, that is, just after explosions of the first stars, here we focus on the astrophysical scenario and investigate the possibility that CRs cause magnetic field generation.
\cite{Miniati2011} proposed that magnetic fields are generated by the resistivity of return current induced by streaming CRs (this mechanism is originally proposed by \cite{Bell2003} in the context of laser-plasma experiments).
If there is temperature inhomogeneity, the resistivity works inhomogeneously and induces rotational electric fields, and thus magnetic fields are generated.
We found that, in the same system, the magnetic field is generated by a different mechanism, that is, the Biermann battery effect.
This driving mechanism of the Biermann battery by CRs through resistive heating has never been discussed.
In this paper, we discuss this mechanism and show that this can be dominant, especially in small-scale structures with $L \lesssim 1 \mathrm{kpc}$.

The rest of this paper is organized as follows.
Section \ref{sec:2} is dedicated to the introduction of equations and the application of this mechanism to the early universe.
We compare this mechanism to other previously proposed ones in Section \ref{sec:3}.
Discussion and summary are appeared in Section \ref{sec:4} and Section \ref{sec:5}, respectively.

\section{Generation mechanism}
\label{sec:2}
\subsection{Equations}
\label{subsec:21}
In order to discuss the generation mechanisms of magnetic fields, first of all, we derive the generalized Ohm's law.
It can be obtained from the weighted sum of equations of motion for each species of plasma fluid and written in the following form.
\begin{align}
    & \pd{}{t} \left( \sum_s q_s n_s \bm{V}_s \right) + \nabla \cdot \left( \sum_s q_s n_s \bm{V}_s \bm{V}_s \right) \nonumber \\
    = & \sum_s \frac{q_s^2 n_s}{m_s} \left( \bm{E} + \frac{\bm{V}_s \times \bm{B}}{c} \right)
    \label{eq:ohm} \\
    & - \sum_s \sum_{s^{\prime} \neq s} \frac{q_s n_s (\bm{V}_s - \bm{V}_{s^{\prime}})}{\tau_{s s^{\prime}}}
    + \sum_s \frac{q_s}{m_s} \left( \bm{f}_s - \nabla P_s \right) \nonumber
\end{align}
$q_s$, $m_s$, $n_s$, $\bm{V}_s$, and $P_s$ are the charge, mass, number density, mean velocity, and pressure of the species $s$, respectively.
The electric and magnetic fields are denoted by $\bm{E}$ and $\bm{B}$.
The second term on the right-hand side represents the effect of collisions and the sum is taken over all pairs of different two species $s$ and $s^{\prime}$.
$\tau_{s s^{\prime}}$ is the time scales of collisions between $s$ and $s^{\prime}$.
$\bm{f}_s$ stands for forces acting on the species $s$ other than electromagnetic, collisional, and pressure gradient forces.
Although the charge and current neutralities can be violated for some instance, electrons rapidly respond in the time scale of plasma oscillation and neutralize the charge and current densities.
Because we are interested in the time scale much larger than the plasma oscillation, the current neutrality condition
\begin{equation}
    \sum_s q_s n_s \bm{V}_s = 0
    \label{eq:current_neutrality}
\end{equation}
can be safely assumed and the first term on the left-hand side vanishes.
%is approximately satisfied and the first term on the left-hand side vanishes.
%The first term on the left-hand side represents the time evolution of total current and this can be approximately regarded as zero because of the current neutrality condition.
The second term on the left-hand side is related to convection.
%The first, second, and third terms on the right-hand side are electromagnetic force, collisions between different species, and pressure gradient and other forces, respectively.

We consider a three-component plasma that consists of thermal electrons ($s={\rm e}$), thermal protons ($s={\rm p}$), and CRs ($s = \mathrm{CR}$).
Then, leaving only the relevant terms for our discussion, the electric field can be expressed as follows.
\begin{equation}
    \bm{E} = - \frac{\bm{V}_{\rm e}}{c} \times \bm{B} + \frac{m_{\rm e}}{e^2 n_{\rm e}} \nabla \cdot \left( \sum_s q_s n_s \bm{V}_s \bm{V}_s \right) - \frac{\nabla P_{\rm e}}{e n_{\rm e}} + \eta \bm{J}_{\rm t}
    \label{eq:electric}
\end{equation}
We neglected some proton terms because the proton mass is much greater than that of electrons. 
The second term on the right-hand side is relevant for the magnetic field generation mechanism discussed in \citet{Ohira2020}, but we neglect this in the rest of this paper.
The last term on the right-hand side emerges from the collision term in equation (\ref{eq:ohm}).
We can neglect the collisions with CRs because, due to the high energy of CRs, the collision time is much longer than that of collisions between thermal protons and thermal electrons.
Then, introducing the resistivity $\eta=m_{\rm e} / n_{\rm e} e^2 \tau_{\rm ep}$ and thermal current $\bm{J}_{\rm t} = \sum_{s \neq {\rm CR}} q_s n_s \bm{V}_s$, the collisional term becomes $\eta \bm{J}_{\rm t}$.
Note that $\bm{J}_{\rm t}$ is nonzero because the current neutrality condition (equation (\ref{eq:current_neutrality})) is satisfied under the presence of the CR current.

Taking the curl of equation (\ref{eq:electric}), we can obtain the time evolution of the magnetic field from Faraday's law.
\begin{align}
    \pd{\bm{B}}{t} & = -c \nabla \times \bm{E} \nonumber \\
    & = \nabla \times (\bm{V}_{\rm e} \times \bm{B}) - \frac{c^2 \eta}{4\pi} \nabla^2 \bm{B} \nonumber \\
    & \qquad -\frac{c}{e n_{\rm e}^2} \nabla n_{\rm e} \times \nabla P_{\rm e} + c \nabla \times(\eta \bm{J}_{\rm CR}).
    \label{eq:generation}
\end{align}
Here, we applied Amp\`ere's law $\nabla \times \bm{B} = (4\pi/c) (\bm{J}_{\rm t} + \bm{J}_{\rm CR})$ for thermal current $\bm{J}_t$.
The CR current density is defined as $\bm{J}_{\rm CR} = q_{\rm CR} n_{\rm CR} \bm{V}_{\rm CR}$.

In these equations, we have not included the effect of the expansion of the universe, even when we apply them to the early universe.
As we see in Section \ref{sec:3}, our mechanism works efficiently in the vicinity of CR sources that are already virialized, and in that case, expansion of the universe can safely be neglected.
If we consider a large cosmological scale, the expansion effect must be included \citep{Naoz2013}.

When collisions in the plasma are dominated by Coulomb collisions, the resistivity can be described by the following so-called Spitzer value.
\begin{equation}
    \eta = 7.23 \times 10^{-9} \log \Lambda \left( \frac{T}{1 \mathrm{K}} \right)^{-3/2} \mathrm{sec} ~,
    \label{eq:Spitzer}
\end{equation}
where $\log \Lambda$ is the Coulomb logarithm and $\log \Lambda = 20$ is assumed in this work.
Because it depends on plasma parameters through logarithmic function, this specific choice of Coulomb logarithm does not affect this story significantly.

\citet{Miniati2011} pointed out that because of the dependence of resistivity on temperature, magnetic fields are generated if there exists temperature inhomogeneity.
This effect is described by the last term on the right-hand side of equation (\ref{eq:generation}).
In addition to this, we discovered that magnetic fields can be generated in the same system but by another mechanism, that is, the Biermann battery effect.
The Biermann battery is a generation mechanism of magnetic fields, where baroclinity of electron density and pressure ($\nabla n_{\rm e} \times \nabla P_{\rm e}$) yields a rotational electron flow \citep{Biermann1950}.
If CRs stream inhomogeneously, Joule heating of the return current raises the plasma temperature inhomogeneously.
Then, the temperature gradient $\nabla T_{\rm e}$ is generated, which is not necessarily parallel to the density gradient $\nabla n_{\rm e}$ because the density structure has already been generated independently of the CR streaming.
Therefore, baroclinicity $\nabla n_{\rm e} \times \nabla P_{\rm e} \neq 0$ is generated by the inhomogeneous CR current and thus the Biermann battery can work even if $\nabla n_{\rm e} \times \nabla P_{\rm e}$ is initially zero.

In the next subsection, by exploiting some assumptions which are also used in \citet{Miniati2011}, we construct simple analytical solutions.

\subsection{Evolution of magnetic fields}
\label{subsec:22}
In this subsection, we derive an analytical solution for the time evolution of magnetic fields following the same procedure as that of \citet{Miniati2011}.
The temperature evolution by the Joule heating induced by CR streaming is described by the following equation:
\begin{equation}
    \frac{\mathrm{d}}{\mathrm{d}t} \left( \frac{3}{2} n k_{\rm B} T \right) = \eta J_{\rm CR}^2,
    \label{eq:evolution_tmp}
\end{equation}
where $n$ is the total number density including both charged and neutral particles since we assume that a temperature equilibrium between charged and neutral particles is established.
This assumption is valid as long as the thermal equilibrium is rapidly achieved compared to the evolution of the magnetic field.
Then the solution of equation (\ref{eq:evolution_tmp}) is
\begin{equation}
    T = T_0 \left( 1 + \frac{t}{\tau} \right)^{2/5},
    \quad \tau = \frac{3nk_{\rm B} T_0}{5\eta_0 J_{\rm CR}^2}.
    \label{eq:temperature}
\end{equation}
We assumed that $n$ and $J_{\rm CR}$ are constant in time and $\eta$ is given by the Spitzer value.
$T_0$ and $\eta_0$ are temperature and resistivity at $t=0$, that is just before the heating.

The magnetic-field generation by the Biermann battery is described by
\begin{equation}
    \frac{\mathrm{d} B}{\mathrm{d} t}
    = - \frac{c}{en_{\rm e}^2} \nabla n_{\rm e} \times \nabla P_{\rm e} = \frac{c k_{\rm B}}{e L^2} T.
    \label{eq:Biermann}
\end{equation}
We assumed that the gradients of electron density and pressure can be simply estimated by using a scale length $L$.
We also assumed thermal equilibrium for electrons and used $P_{\rm e} = n_{\rm e} k_{\rm B} T$.
equation (\ref{eq:Biermann}) can be solved easily:
\begin{equation}
    B = \frac{3}{7} B_* \zeta^2 \left[ \left( 1 + \frac{t}{\tau} \right)^{7/5} - 1 \right].
    \label{eq:new}
\end{equation}
Here we defined $B_*$ and $\zeta$ as
\begin{equation}
    B_* = \eta_0 n e c, \quad
    \zeta = \frac{k_{\rm B}T_0}{e \eta_0 J_{\rm CR} L}.
    \label{eq:zeta}
\end{equation}
We estimate the achievable field strength for the environment in the early universe in the next subsection.

\subsection{Application to the early universe}
\label{subsec:23}
\citet{Miniati2011} estimated the CR flux at a distance $R$ from a galaxy as follows.
\begin{equation}
    n_{\rm CR} V_{\rm CR} \simeq 33  \left( \frac{\mathcal{L}}{\mathcal{L}_*} \right) \left( \frac{R}{1 \, \mathrm{kpc}} \right)^{-2} \, \mathrm{cm^2 \, s^{-1}},
    \label{eq:current}
\end{equation}
where $\mathcal{L}_* \simeq 5.2 \times 10^{28} \, \mathrm{erg \, s^{-1} \, Hz^{-1}}$ is the typical luminosity of bright galaxies at redshift $z \geq 6$.
Then, in the early universe, parameters appeared in equation (\ref{eq:new}) take the following values:
\begin{align}
    \tau & \simeq 2.5 \times 10^3 \, \mathrm{sec} \, \left( \frac{n}{10^{-4} \, \mathrm{cm^{-3}}} \right) \nonumber \\
    & \qquad \cdot \left( \frac{T_0}{1 \, \mathrm{K}} \right)^{5/2} \cdot \left( \frac{n_{\rm CR} V_{\rm CR}}{10 \, \mathrm{cm^{-2} s^{-1}}} \right)^{-2},
\end{align}
\begin{equation}
    B_* \simeq 2.1 \times 10^{-10} \, \mathrm{G} \, \left( \frac{T_0}{1 \, \mathrm{K}} \right)^{-3/2} \left( \frac{n}{10^{-4} \, \mathrm{cm^{-3}}} \right),
\end{equation}
\begin{align}
    \zeta & \simeq 1.4 \times 10^{-13} \left( \frac{T_0}{1 \, \mathrm{K}} \right)^{5/2} \nonumber \\
    & \qquad \cdot \left( \frac{n_{\rm CR} V_{\rm CR}}{10 \, \mathrm{cm^{-2} s^{-1}}} \right)^{-1} \cdot \left( \frac{L}{1 \, \mathrm{kpc}} \right)^{-1}.
\end{align}
Then, the field strength at $t \ll \tau$ is estimated as
\begin{align}
    B & \simeq 9.9 \times 10^{-40} \, \mathrm{G} \, \left( \frac{T_0}{1 \, \mathrm{K}} \right) \nonumber \\
    & \qquad \cdot \left( \frac{n}{10^{-4} \, \mathrm{cm^{-3}}} \right)
    \cdot \left( \frac{L}{1 \, \mathrm{kpc}} \right)^{-2} \cdot \left( \frac{t}{1 \, \mathrm{sec}} \right).
\end{align}
For $t \gg \tau$, it is
\begin{align}
    B & \simeq 3.9 \times 10^{-18} \, \mathrm{G} \, \left( \frac{n}{10^{-4} \, \mathrm{cm^{-3}}} \right)^{-2/5} \nonumber \\
    & \qquad \cdot \left( \frac{L}{1 \, \mathrm{kpc}} \right)^{-2} \cdot \left( \frac{n_{\rm CR} V_{\rm CR}}{10 \, \mathrm{cm^{-2} s^{-1}}} \right)^{4/5} \cdot \left( \frac{t}{1 \, \mathrm{Gyr}} \right)^{7/5}.
    \label{eq:strength}
\end{align}
This field strength is sufficient for the seed field that will be amplified to the $\mathrm{\mu G}$ level over cosmological time scales \citep{Davis1999}.

\section{Comparison with other mechanisms}
\label{sec:3}
A few other mechanisms that generate magnetic fields by CRs have been proposed so far.
One is the resistive mechanism given by \citet{Miniati2011}.
\citet{Ohira2021} discussed that the Biermann battery is induced by the electron flow driven in response to the streaming CRs.
Here, we compare these mechanisms to the mechanism proposed in this work.

Time evolution of $B$ by the mechanism of \citet{Miniati2011} can be expressed by
\begin{equation}
    \frac{\mathrm{d} B_{\rm MB11}}{\mathrm{d} t}
    = \frac{c \eta_0 J_{\rm CR}}{L} \left( \frac{T}{T_0}\right)^{-3/2}.
    \label{eq:resistive}
\end{equation}
In our notation, the solution can be written in the following form:
\begin{equation}
    B_{\rm MB11} = \frac{3}{2} B_* \zeta \left[ \left( 1 + \frac{t}{\tau} \right)^{2/5} - 1 \right],
    \label{eq:miniati}
\end{equation}
where the same assumptions used to derive equation (\ref{eq:new}) are made.

In the same way, the time evolution of the magnetic field by \citet{Ohira2021}'s mechanism can be obtained analytically.
The time evolution of the electron pressure can be described by the energy equation of fluid dynamics,
\begin{equation}
    \pd{P_{\rm e}}{t} + (\bm{V}_{\rm e} \cdot \nabla) P_{\rm e} = - \gamma \nabla \cdot \bm{V}_{\rm e}.
    \label{eq:evolution_p}
\end{equation}
$\gamma$ is the specific heat ratio of the electron fluid.
Assuming that the advection of electron pressure is negligible and the electron flow is constant in time, this can be solved to yield the following solution:
\begin{equation}
    P_{\rm e} = P_{{\rm e}0} \exp( -\gamma t\nabla \cdot \bm{V}_{\rm e} ), %= \exp \left( \frac{\gamma J_{\rm CR}}{en_{\rm e} L} t \right),
    \label{eq:pressure}
\end{equation}
where $P_{{\rm e}0} = n_{\rm e} k_{\rm B} T_0$ is the initial electron pressure.
The evolution of the magnetic field is again determined by equation (\ref{eq:Biermann}) but with $P_{\rm e}$ given by (\ref{eq:pressure}).
\begin{equation}
    \od{B}{t} = \frac{P_{\rm e} \gamma c t}{e n_{\rm e}^2} \nabla n_{\rm e} \times \nabla (\nabla \cdot \bm{V}_{\rm e})
\end{equation}
Using the current neutrality condition $n_{\rm e} \bm{V}_{\rm e} = n_p \bm{V}_p + \bm{J}_{\rm CR}/e$ and $\nabla \to 1/L$ and assuming that $\gamma t \nabla \cdot \bm{V}_{\rm e} \ll 1$ and therefore $P_{\rm e}$ is almost constant, this equation can be solved analytically.
The solution in our notation is
\begin{equation}
    B_{\rm O21} = \frac{9\gamma}{25 \chi_{\rm e}} B_* \zeta^3 \left( \frac{t}{\tau} \right)^2,
    \label{eq:ohira}
\end{equation}
where $\chi_{\rm e}=n_{\rm e}/n$ is the ionization fraction.
This mechanism works more efficiently for a weakly ionized plasma because a faster electron flow is needed to neutralize the incoming CR current and therefore the electron fluid is more rapidly compressed.
Although we neglected the Joule heating in equation (\ref{eq:evolution_p}), we note a possibility where the Joule heating and adiabatic compression induced by the electron flow work at the same time.
In that case, the pressure is raised by both the effects and stronger magnetic fields can be generated.
In the following, however, we investigate the dominance of these mechanisms assuming that they work independently.

\begin{figure}
    \centering
    \includegraphics[width=80mm]{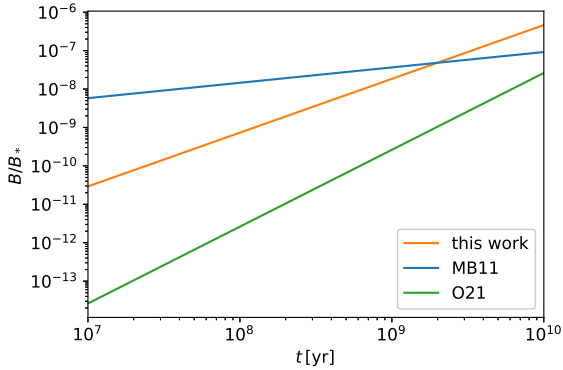}
    \caption{Time evolution of the magnetic field by different mechanisms.
    The orange line corresponds to the mechanism proposed in this work.
    The blue and green lines show those by \citet{Miniati2011} and \citet{Ohira2021}, respectively.}
    \label{fig:evolution}
\end{figure}

Fig. \ref{fig:evolution} is the time evolution of each mechanism given by analytical solutions (\ref{eq:new}), (\ref{eq:miniati}), and (\ref{eq:ohira}) for fiducial parameters (see subsection \ref{subsec:23}) and $\chi_{\rm e} = 10^{-3}$.
For simplicity, we assumed that all the parameters that appeared in the analytical solutions, including $n$ and $\chi_{\rm e}$, are independent of time, although they can vary in time in realistic situations.
For these parameters, our new mechanism overtakes that of \citet{Miniati2011} in $t\simeq 2 \times 10^9 \, \mathrm{yr}$, but \citet{Ohira2021}'s mechanism remains less efficient than the other two.

\begin{figure*}
    \centering
    \includegraphics[width=160mm]{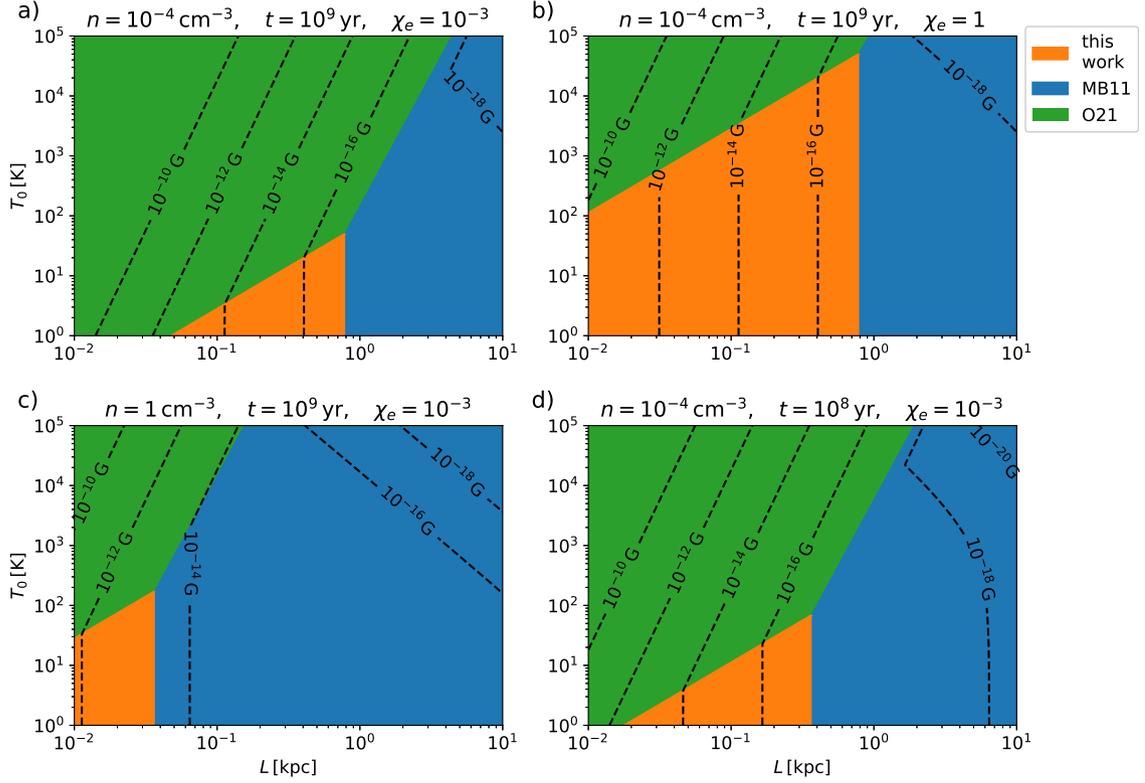}
    \caption{
    Parameter space where different generation mechanisms dominate.
    $L$ is the length scale and $T_0$ denotes the initial temperature.
    The orange, blue, and green regions are where the generation mechanism by this work, \citet{Miniati2011} and \citet{Ohira2021} dominate, respectively.
    The black dotted contours indicate the estimated field strength achieved by the dominant generation mechanism.
    The CR current is varied as $n_{\rm CR} V_{\rm CR} =  10 \, (L/ 1 \, \mathrm{kpc})^{-2} \, \mathrm{cm^{-2} \, s^{-1}}$.
    The values of the other parameters ($n$, $t$, and $\chi_{\rm e}$) are written at the top of each panel a) - d).
    }
    \label{fig:allmap}
\end{figure*}

Next, we investigate the parameter dependence of these mechanisms.
Although there are many uncertain parameters, first we check the dependence on scale-length $L$ and initial temperature $T_0$ with $n=10^{-4} \, \mathrm{cm^{-3}}$ and $\chi_{\rm e} = 10^{-3}$ fixed.
Considering the length scale $L$ which is equal to the distance from a galaxy $R$ in equation (\ref{eq:current}), the CR current is also varied as a function of $L$, that is $n_{\rm CR} V_{\rm CR} =  10 \, (L/ 1 \, \mathrm{kpc})^{-2} \, \mathrm{cm^{-2} \, s^{-1}}$.
The panel a) of Fig. \ref{fig:allmap} shows which of the three mechanisms is dominant at $t = 10^9 \mathrm{yr}$.
The orange, blue, and green regions correspond to the regions where the mechanism in this work, that of \citet{Miniati2011}, and that of \citet{Ohira2021} dominate, respectively. 
We overplotted the estimated strength of magnetic fields generated by the corresponding dominant mechanism.

Next, we investigate the fully-ionized case ($\chi_{\rm e}=1$), high-density case ($n=1 \, \mathrm{cm^{-3}}$), and short-time case ($t=10^8 \, \mathrm{yr}$) in the panel b), c), and d) in Fig. \ref{fig:allmap}, respectively.
All the parameters other than that changed are fixed to those for the panel a).
These panels show that our new mechanism is efficient in the region which has a small scale length and low initial temperature, while for a larger scale, \citet{Miniati2011}'s mechanism dominates.
The mechanism proposed by \citet{Ohira2021} is efficient in the high temperature and/or less ionized region.

\section{Discussion}
\label{sec:4}
As clarified in the previous section, the dominance of the generation mechanisms of magnetic fields by CRs depends on the plasma environment including number density $n$, ionization fraction $\chi_{\rm e}$, and temperature $T$.
When CRs propagate through a plasma, CRs themselves also alternate the plasma environment.
Although the temperature variation by the Joule heating of return current induced by CR streaming is explicitly described in equation (\ref{eq:evolution_tmp}), ionization by CRs and heating by the ionization and Coulomb interactions with CRs are not included.
\cite{Leite2017} analyzed the role of CRs on ionization and heating of the intergalactic medium (IGM) and showed that CRs can substantially heat the IGM raising its temperature by $\Delta T = 10 - 200 \mathrm{K}$ before the epoch of reionization, while they hardly contribute to the ionization.
Here we evaluate these effects and compare them with the Joule heating of the return current.

The ionization rate and heating rate by ionization and Coulomb interactions with CRs are given in \citet{Jasche2007}.
The ionization rate $\Gamma_{\rm ion}^{\rm CR}$ is
\begin{equation}
    \Gamma_{\rm ion}^{\rm CR} = \frac{1}{n W_{\rm H}} \int_{E_{\rm min}}^{\infty} \left| \frac{\mathrm{d} E}{\mathrm{d} t} \right|_{I} \frac{\mathrm{d} n_{\rm CR}}{\mathrm{d} E} \, \mathrm{d} E,
    \label{eq:ionization_rate}
\end{equation}
where $W_H \simeq 36.3 \, \mathrm{eV}$ is the mean energy expended by a CR proton to create an ion pair.
For a neutral medium ($\chi_{\rm e}=0$), $\xi$ takes $5/3$.
The total heating rate $\mathcal{H}^{\rm CR}$ is given by
\begin{equation}
    \mathcal{H}^{\rm CR} = [W_H - \xi I_{\rm H}] \Gamma_{\rm ion}^{\rm CR} + \mathcal{H}_{\rm Coulomb},
    \label{eq:CR_heating}
\end{equation}
where $I_H = 13.6 \, \mathrm{eV}$ is the ionization potential and $\mathcal{H}_{\rm Coulomb}$ is the heating rate by Coulomb interaction.
$\xi$ is the factor to correct the contribution from all secondary and higher generation ionization.
For a weakly ionized plasma ($\chi_{\rm e} \ll 1$) as encountered in the early universe, Coulomb interaction is less efficient than ionization.
Therefore, the contribution from the second term in equation (\ref{eq:CR_heating}) is negligible.
For further simplification, we assume that the plasma is composed only of protons and the CRs have the energy distribution of delta function centered at $E = 1 \, \mathrm{GeV}$.
Then, the heating rate per unit volume is evaluated as follows.
\begin{align}
    n \mathcal{H}^{\rm CR} & = 6.7 \times 10^{-33} \, \mathrm{erg \cdot cm^{-3} \cdot s^{-1}} \nonumber \\
    & \qquad \quad \cdot \left( \frac{n}{10^{-4} \, \mathrm{cm^{-3}}} \right)  \cdot \left( \frac{n_{\rm CR} V_{\rm CR}}{10 \, \mathrm{cm^{-2} s^{-1}}} \right).
    \label{eq:ionization_heating}
\end{align}
On the other hand, the heating rate by the Joule heating is
\begin{align}
    \eta J_{\rm CR}^2 & = 3.4 \times 10^{-24} \, \mathrm{erg \cdot cm^{-3} \cdot s^{-1}} \nonumber \\
    & \qquad \quad \cdot \left( \frac{T}{1 \, \mathrm{K}} \right)^{-3/2} \cdot \left( \frac{n_{\rm CR} V_{\rm CR}}{10 \, \mathrm{cm^{-2} s^{-1}}} \right)^2.
    \label{eq:Joule_heating}
\end{align}
The use of equation (\ref{eq:evolution_tmp}) is safely justified for the application to the early universe since the Joule heating is more efficient than the heating by ionization for $T < 6.4 \times 10^5 \, \mathrm{K} \, (n/10^{-4}\, \mathrm{cm^{-3}})^{-2/3} \cdot (n_{\rm CR} V_{\rm CR}/10 \, \mathrm{cm^{-2} s^{-1}})^{2/3}$.
The other point these estimates imply is that the Joule heating caused by the streaming CRs is more important for the heating in the early universe.
This effect has never been discussed and should be examined in  future works.

In addition to the Joule heating and heating by ionization, we have to consider cooling processes to determine the plasma temperature.
When the cooling processes work efficiently, the Biermann battery effect is suppressed and the generation mechanism by \citet{Miniati2011} tends to dominate.
The cooling rate depends on the plasma environment including its composition and density and it greatly changes before and after the reionization.
Therefore, the generation of the magnetic field must be solved with the evolution of the plasma in cosmological simulations, taking all the heating and cooling mechanisms into account.
This should also be done in future studies.

In the previous sections, we presumed that the gradients of temperature and density are not parallel.
As implied by equation (\ref{eq:evolution_tmp}), this requirement is expected to be satisfied where CRs stream inhomogeneously, because the direction of temperature gradient produced by the Joule heating does not necessarily coincide with that of density gradient which is set before the heating.
The situation considered in Section \ref{subsec:23} is a good example of inhomogeneous $J_{\rm CR}$ because, as is evident from equation (\ref{eq:current}), the CR current density from a galaxy varies as a function of radial distance $R$.
Even if CRs do not propagate inhomogeneously, there are some situations where misalignment of the density and temperature gradients can be produced.
Shocks in forming galaxies and supernova remnants are examples of such situations \citep{Kulsrud1997, Hanayama2005}.
Because the evolution of the density structures is nonlinear in those objects, numerical simulations are necessary to confirm if the structures preferable for the Biermann mechanism are produced.
On the other hand, when we consider a larger system where the cosmic expansion works, we can follow the evolution of magnetic fields by using the linear theory of the density evolution \citep{Naoz2013}.
Whether or not inconsistency between the gradients of temperature and density can be realized is an important problem but beyond the scope of this paper.
The strength of CR current is another uncertainty and also an important problem.
These points should be investigated further in realistic situations.

The resistive term in Equation (\ref{eq:generation}) also contributes to the dissipation of the generated magnetic fields.
We can estimate the time scale of dissipation $\tau_{\rm dis}$ by
\begin{equation}
    \tau_{\rm dis} = \frac{4\pi L^2}{c^2 \eta}
    = 2.9 \times 10^{13} \, \mathrm{Gyr}
    \left( \frac{L}{1 \, \mathrm{kpc}} \right)^2
    \left( \frac{T}{1 \, \mathrm{K}} \right)^{3/2}.
\end{equation}
Because the dissipation time is much longer than the timescale that we consider, the dissipation is negligible in this work.
In the same way, we can evaluate the effect of magnetic reconnection by its time scale $\tau_{\rm rec} = L/v_{\rm A}$, where $v_{\rm A}=B/\sqrt{4\pi m_{\rm p} n_{\rm p}}$ is the Alfv\'en velocity:
\begin{equation}
    \tau_{\rm rec}% = \frac{L}{v_A}
    = 1.4 \times 10^{10} \, \mathrm{Gyr}
    \left( \frac{L}{1 \, \mathrm{kpc}} \right)
    \left( \frac{B}{10^{-20} \, \mathrm{G}} \right)^{-1}
    \left( \frac{n_{\rm p}}{10^{-7} \, \mathrm{cm^{-3}}} \right)^{1/2}
\end{equation}
Because this is much longer than the evolution time scales of magnetic fields and the universe, large-scale magnetic fields are not subject to magnetic reconnection.

Finally, we would like to mention a self-discharge by the streaming CRs \citep{Ohira2022}.
As shown in equation (\ref{eq:electric}), the electric field is induced by streaming CRs.
Then, secondary electrons produced by CR ionization could be accelerated by the electric field.
Then, the return current of thermal electrons is replaced by the electric current of secondary electrons.
Since the secondary electrons have large energies compared with the thermal electrons, the resistivity of the secondary electrons is much smaller than one of the thermal electrons. 
As a result, the resistive electric field becomes small, resulting in suppression of the magnetic field generation proposed by \citet{Miniati2011}.
Although the other two generation mechanisms of the magnetic field should be also affected by the self-discharge, the Biermann battery can work as long as baroclinicity $\nabla n_{\rm e} \times \nabla P_{\rm e}$ does not disappear. 
Therefore, we need further studies to understand the magnetic field generation by CRs.

\section{Summary}
\label{sec:5}
We discovered a new driving mechanism of the Biermann battery, that is, resistive heating induced by streaming CRs.
The expected field strength generated by this mechanism is sufficiently strong to be amplified to the currently observed $\mathrm{\mu G}$ level.
The comparison with other previously proposed mechanisms (Fig. \ref{fig:allmap}) revealed that the mechanism of this work can dominate where plasma temperature is relatively low, the scale length is small, and the ionization fraction is large.

\section*{acknowledgments}
This work is supported by JSPS KAKENHI grant numbers JP21J20737 (SY), JP19H01893 (YO), and JP21H04487 (YO).
YO is supported by Leading Initiative for Excellent Young Researchers, MEXT, Japan.

\section*{Data availability}
No new data were generated or analysed in support of this research.

\bibliographystyle{mnras}
\bibliography{library}

\bsp	% typesetting comment
\label{lastpage}
\end{document}